# Planet migration in massive circumbinary discs

Matthew Teasdale ⋆ and Dimitris Stamatellos ⋆

*Jeremiah Horrocks Institute for Mathematics, Physics and Astronomy, University of Central Lancashire, Preston PR1 2HE, UK*

## ABSTRACT

Most stars are in multiple systems, with the majority of those being binaries. A large number of planets have been confirmed in binary stars and, therefore, it is important to understand their formation and dynamical evolution. We perform simulations to investigate the migration of wide-orbit giant planets (semimajor axis 100 au) in massive circumbinary discs (mass $0.1\,M_\odot$) that are marginally gravitationally unstable, using the three-dimensional Smoothed Particle Hydrodynamic code SEREN. We vary the binary parameters to explore their effect on planet migration. We find that a planet in a massive circumbinary disc initially undergoes a period of rapid inward migration before switching to a slow outward migration, as it does in a circumstellar disc. However, the presence of the binary enhances planet migration and mass growth. We find that a high binary mass ratio (binary with equal mass stars) results in more enhanced outward planet migration. Additionally, larger binary separation and/or higher binary eccentricity results to a faster outward planet migration and stronger planet growth. We conclude that wide-orbit giant planets attain wider final orbits due to migration around binary stars than around single stars.

**Key words:** accretion, accretion discs – hydrodynamics – radiative transfer – protoplanetary discs – (stars:) binaries: general.

## 1 INTRODUCTION

Since the discovery of the first exoplanet around a main-sequence star, 51 Pegasi b (Mayor & Queloz 1995), more than 5300 exoplanets have been confirmed.[1] Many of these exoplanets have been observed orbiting binaries and come in two types; S or satellite-type, in which the planet orbits around one star of the binary, and P or planetary-type planet, in which the planet orbits both stars of the binary (Dvorak 1984). The first P-type, i.e. circumbinary, planet detected is Kepler-16b (Doyle et al. 2011) and since then 28 circumbinary planets have been confirmed (Su et al. 2021; http://exoplanet.eu).

It is believed that around half of stars are in multistellar systems. Raghavan et al. (2010) studied a sample of 454 stars selected from the Hipparcos catalogue in the solar neighbourhood. They found that 44 per cent stars are within multistar systems with 150 of the 454 sample stars in binary systems. It is, therefore, important to understand how planets form and evolve in such multiple systems (see review by Marzari & Thebault 2019), and whether there are any distinct differences from the formation and evolution of planets around single stars.

Currently there are two widely accepted theories for the formation of gas giant planets: (i) core accretion, and (ii) gravitational instability.

The core accretion model is thought to account for the origin of the majority of giant exoplanets currently known. This theory suggests that a core forms through the accretion of both pebbles and planetesimals within a gaseous disc (Goldreich & Ward 1973; Mizuno 1980; Bodenheimer & Pollack 1986; Pollack et al. 1996; Drazkowska et al. 2023). This accretion of material may lead to the core having a sufficient mass to attain a gaseous envelope. The growth of the massive core, $\sim 10\,$Myr (Pollack et al. 1996), is at odds with the dispersal of the protoplanetary disc (3–5 Myr) and so core accretion has difficulty in forming gas giants on wide orbits (Wagner, Apai & Kratter 2019). In the core accretion model, Jupiters are expected to form outside the snowline ($\sim 3\,$au for solar-type stars). In this region, dust particles have an ice coating which promotes fast dust growth, as dust particles are more 'sticky'. Within the snow line, dust particles cannot coagulate as efficiently and, therefore, only terrestrial planets may form (Drazkowska et al. 2023). One potential way of explaining close-in giant planets (hot Jupiters) is by assuming that a planet migrates to its final location from farther out in the disc (Dawson & Johnson 2018).

⋆ E-mail: MTeasdale1@uclan.ac.uk (MT); Dstamatellos@uclan.ac.uk (DS)

[1]NASA Exoplanet Archive, DOI: 10.26133/NEA12 (Accessed on: 05/04/2022).

Gas giant planets may also form through gravitational fragmentation of discs. A disc becomes gravitationally unstable if it is sufficiently cold and/or massive so that the Toomre criterion is satisfied (Toomre 1964),

$$Q \equiv \frac{c_s \Omega}{\pi G \Sigma} \lesssim Q_{crit} \simeq 1-2,\tag{1}$$

where $Q$ is the Toomre parameter, $G$ is the gravitational constant, $c_s$ is the sound speed, $\Omega$ is the angular frequency, and $\Sigma$ the surface density of the disc. One of the outcomes of gravitational instability is fragmentation which can lead to the formation of giant planets. Gammie (2001) shows that if the cooling time of the disc is sufficiently short, $\tau_c \lesssim 3\Omega^{-1}$, then the disc can fragment. The above conditions can be satisfied at large radii where; therefore, fragmentation is likely. It is expected that gravitational instability commonly produces massive planets at large separations (Matzner & Levin 2005; Whitworth & Stamatellos 2006; Stamatellos, Hubber & Whitworth 2007a; Clarke & Lodato 2009; Stamatellos & Whitworth 2009a, b; Kratter, Murray-Clay & Youdin 2010; Mercer & Stamatellos 2017, 2020). However, observations show that such planets are not very common (e.g. Vigan et al. 2021; Rice 2022).

Once a planet forms in a disc, it interacts with it and then may migrate inwards or outwards, depending on the balance between the torques that are exerted from the inner and outer disc onto the planet. Planetary migration can be described by two general types: Type I and Type II. Type I migration happens when the gravitational interaction between a planet and the disc does not significantly alter the structure of the disc (Ward 1997; Tanaka, Takeuchi & Ward 2002). The planet drifts relative to the disc on a time-scale that is inversely proportional to the mass of the planet. Type I migration relates to low-mass planets (e.g. Earth-mass planets) that are not able to open a gap in the disc. Type II migration happens when the planet is massive enough to open a gap in the disc so that a flow barrier to the disc gas may be established (Ward 1997; Paardekooper et al. 2022). The planet becomes locked into the disc as it acts as a 'bridge' to transport angular momentum from the inner region of the disc to the outer disc region (i.e. outside the gap), in effect compensating for the lack of gas in this region. As such the disc evolves viscously as it would have done without the planet. As the disc evolves, the planet migrates inwards on a time-scale set by the disc's viscosity. The standard Type II migration is inwards. Kanagawa, Tanaka & Szuszkiewicz (2018) propose a new picture of Type II migration in which the planet migrates with Type I-like torque. This is different to the classical view of Type II migration as they argue that even the gap has low density, Type I torques are still important and need to be taken into account (Paardekooper et al. 2022). This new model can be used to describe the transition from Type I to Type II migration. Migration may also happen in an outward direction under certain conditions (Bitsch & Kley 2010; Lin & Papaloizou 2012; Cloutier & Lin 2013; Lega et al. 2015; Stamatellos 2015; Stamatellos & Inutsuka 2018; Dempsey, Muñoz & Lithwick 2021).

Planets that form on wide orbits early-on during the disc evolution (e.g. by disc instability) are expected to interact with the gas-rich early-phase disc and migrate. Baruteau, Meru & Paardekooper (2011) suggests that newly formed, Jupiter-sized planets will undergo Type I migration and possibly get destroyed as they reach the host star. However, other studies argue that such planets can open up a gap in the disc, so that their migration slows down (Zhu et al. 2012; Stamatellos 2015; Stamatellos & Inutsuka 2018; Vorobyov & Elbakyan 2018). Stamatellos (2015) and Stamatellos & Inutsuka (2018) show that a giant planet that forms in massive, marginally gravitationally stable disc, initially migrates rapidly inwards until it is able to open a gap in the disc. Then migration slows down and switches to an outward direction due to the interaction of the planet with the gravitationally unstable edges of the gap (Lin & Papaloizou 2012; Cloutier & Lin 2013). This outwards motion is due to accretion of high angular momentum gas from outside of the planet's orbit. We label this type of migration, in which the planet opens up a gap and migrates outwards, as *non-standard* Type II migration.

The migration of planets in circumbinary discs has been less studied (Pierens & Nelson 2008; Kley & Haghighipour 2015; Mutter, Pierens & Nelson 2017; Thun & Kley 2018; Kley, Thun & Penzlin 2019; Marzari & Thebault 2019). Kley, Thun & Penzlin (2019) study Saturn-mass planets on close orbits (∼0.5 au) within a ∼0.01 $M_\odot$ disc around a tight binary (∼0.2 au) and find that planets migrate inwards towards the inner gap around the binary, in some cases after a brief period of outward migration. On the other hand, Kley & Haghighipour (2015) find that the planet can migrate outwards due to interactions with a highly eccentric disc that has formed as a result of the binary-disc interaction.

The aim of this work is to study the migration of gas giant planets formed on wide orbits in massive circumbinary discs. More specifically, we want to understand how the evolution of the planetary orbit is different in a circumbinary than in a circumstellar disc. We are interested in whether a planet is able to migrate outwards and survive at a wide-orbit in a massive circumbinary disc as it does in a circumstellar disc. We expand upon the work of Stamatellos (2015) and Stamatellos & Inutsuka (2018), with a focus on discs around binary stars. We investigate the effect of the binary mass ratio, binary separation, and binary eccentricity on the migration of a circumbinary gas giant planet that is initially on a wide orbit.

We describe the computational method in Section 2, and in Section 3 the simulation set up. In Section 4, we present the set of simulations performed. In Section 5, we discuss the wider implications of this work before finally concluding in Section 6.

## 2 COMPUTATIONAL METHOD

We simulate the dynamics of a giant planets in a gaseous circumbinary disc using SEREN, a Smoothed Particle Hydrodynamics (SPH) code developed by Hubber et al. (2011). SEREN uses a second-order Runge–Kutta integration scheme and time independent artificial viscosity with $\alpha_{min} = 0.1$, $\alpha = 1$, and $\beta = 2$. Simulations are run using the method of radiative transfer of Stamatellos et al. (2007b). This method uses the gravitational potential, temperature, and density of each SPH particle to estimate its mean optical depth. This regulates its heating and cooling, which is given by

$$\frac{\mathrm{d}u_i}{\mathrm{d}t}\bigg|_{\mathrm{RAD}} = \frac{4\sigma_{\mathrm{SB}}\left(T_A^4 - T_i^4\right)}{\overline{\Sigma}_i^2 \,\overline{\kappa}_{\mathrm{R}}\left(\rho_i, T_i\right) + \kappa_{\mathrm{p}}^{-1}\left(\rho_i, T_i\right)}, \tag{2}$$

where $u_i$ is the specific internal energy of the particle, $\sigma_{\mathrm{SB}}$ is the Stefan–Boltzmann constant, $\overline{\Sigma}$ is the mass-weighted mean column density, $\rho_i$ is the density of the particle, $T_i$ is the temperature of the particle, $\overline{\kappa}_{\mathrm{R}}$ and $\kappa_{\mathrm{p}}$ are the Rosseland- and Planck-mean opacities. $T_A$ is the pseudo-background radiation temperature set to 10 K (see Stamatellos et al. 2007b, for details).

The binary and planet are represented by sink particles and only interact with the rest of the computational domain through their gravity (Bate, Bonnell & Price 1995). The sink radius of the stars and planet is chosen as $R_{\mathrm{sink},\star} = 0.2$ au and $R_{\mathrm{sink,p}} = 0.1$ au, respectively. The sink radius of the planet is chosen so that it is smaller than its Hill radius, which defines the region in which the gravity of the planet dominates the gravity of binary,

$$R_{\mathrm{H}} = \left(\frac{M_{\mathrm{p}}}{3M_\star}\right)^{\frac{1}{3}} a, \tag{3}$$

where $M_{\mathrm{p}}$ is the mass of the planet, $M_\star$ the total mass of the binary, and $a$ the orbital radius of the planet. For a planet of mass $M_{\mathrm{p}} = 1\,M_{\mathrm{J}}$ that orbits a solar-mass star at $a = 50$ au, the Hill radius is $R_{\mathrm{H}} = 3.4$ au, which is much larger than the planet sink radius. This value changes as the planet accretes material from the disc and migrates over time, but it remains larger than the planet's sink radius (see Section 4.3.4).

**Table 1.** The binary parameters of the 10 simulations presented in this paper. $q_b$ is the binary mass ratio, $\alpha_b$ is the binary separation, and $e_b$ is the binary eccentricity.

| Run | $q_b$ | $\alpha_b$ (au) | $e_b$ |
|-----|-------|------------------|-------|
| 1   | 1     | 1                | 0.0   |
| 2   | 1     | 1                | 0.2   |
| 3   | 1     | 5                | 0.0   |
| 4   | 1     | 5                | 0.2   |
| 5   | 1     | 5                | 0.5   |
| 6   | 0.25  | 1                | 0.0   |
| 7   | 0.25  | 1                | 0.2   |
| 8   | 0.25  | 5                | 0.0   |
| 9   | 0.25  | 5                | 0.2   |
| 10  | 0.25  | 5                | 0.5   |

## 3 SIMULATION SET UP

We perform a set of 10 simulations of a planet within a circumbinary disc, and one simulation of a planet in a circumstellar disc as a benchmark simulation. In the 10 runs presented here, we investigate how the initial binary parameters (binary mass, separation, and eccentricity) affect the migration of the planet. The specific parameters that we vary can be seen in Table 1.

### 3.1 Initial conditions

We assume that the circumbinary disc initially extends from $R_{in}^D = 0.1$ au to $R_{out}^D = 100$ au and is of initial mass $M_D = 0.1\,M_\odot$. The disc is represented by $5 \times 10^5$ SPH particles.

We model two initial binary mass ratios, $q_b = 1$ and $q_b = 0.5$. For the initial binary mass ratio $q_b = 1$, the binary components are set to $M_1 = 0.5\,M_\odot$ and $M_2 = 0.5\,M_\odot$, whereas for $q_b = 0.25$, are set to $M_1 = 0.8\,M_\odot$ and $M_2 = 0.2\,M_\odot$. Therefore, in both sets of simulations the total mass of the binary is $M_\star = 1\,M_\odot$. We model two initial binary separations, $\alpha_b = 1, 5$ au, and three initial binary eccentricities, $e_b = 0.0, 0.2, 0.5$.

The surface density profile of the disc is set to

$$\Sigma_0(R) = \Sigma(1\,\mathrm{au})\left(\frac{R}{\mathrm{au}}\right)^{-1}, \tag{4}$$

and the disc temperature profile to

$$T_0(R) = 250\,\mathrm{K}\left(\frac{R}{\mathrm{au}}\right)^{-0.5} + 10\,\mathrm{K}, \tag{5}$$

where $\Sigma(1\,\mathrm{au})$ is determined by the mass and radius of the disc, and $R$ is the distance from the centre of mass of the binary. The initial surface density and temperature profiles, and the resulting $Q$ parameter, for a selection of runs are seen in Fig. 3. The disc is initially within the marginally unstable regime ($Q \sim 1$–2) when $R \gtrsim 30$ au.

The disc is allowed to relax, i.e. evolve without embedding the planet within it, for 3 kyr. The planet has initial mass $M_p = 1\,M_J$, and semimajor axis $\alpha_p = 50$ au. The orbit of the planet is initially circular. The disc–planet system is then allowed to evolve for 20 kyr.

Each of these circumbinary planetary systems have the same total stellar-mass, the same planet properties and the same disc properties to that of the circumstellar disc system modelled by Stamatellos & Inutsuka (2018), so that direct comparisons can be made. We will examine how replacing the single star with a binary of the same total mass will affect the evolution of the planet.

### 3.2 Benchmark run: planet migration in a circumstellar disc

A planet of mass $M_p = 1\,M_J$ is embedded in a circumstellar disc, of mass $M_D = 0.1\,M_\odot$, at $\alpha = 50$ au. This disc is around a star of mass $M_\star = 1\,M_\odot$ and is allowed to relax for 3 kyr. To validate the number of particles that we use, we perform a convergence test increasing the number of SPH particles from $0.5 \times 10^5$ to $6 \times 10^5$, and we find convergence above $4 \times 10^5$ particles. Therefore, for computational efficiency, we chose to use $5 \times 10^5$ SPH particles to represent the disc.

Using this number of SPH particles, we find that the planet's semimajor axis, mass, and eccentricity in this circumstellar system evolves in a way that is very similar to that in Stamatellos & Inutsuka (2018), which have the same initial conditions but use $10^6$ SPH particles to represent the disc. There is a small difference in the evolution of the semimajor axis of the planet, which we attribute to the difference in the number of SPH particles used (resulting in a slightly different artificial viscosity). We will use the results of this run to compare the results of the planet's evolution in a circumbinary disc.

## 4 EVOLUTION OF A PLANET IN A CIRCUMBINARY DISC

The evolution of the giant planet in two representative runs (Run 1 and Run 6) can be seen in Figs 1 and 2, in which the surface density of the disc is plotted at various times. In all runs, the planet initially migrates quickly inwards and grows in mass through accretion from the disc. After ~2.5 kyr, the planet opens a gap in the disc and migration slows down. Following this, the migration reverses direction, and the planet moves outwards.

We will first briefly discuss the evolution of the properties of the disc and the binary before focusing on how the planet migrates in the disc.

### 4.1 Disc evolution: Toomre parameter, surface density and temperature of the disc

The binary and the planet interact with the disc, affecting its structure. The Toomre parameter $Q$, the surface density and the temperature profile for two representative runs (Run 1 and 6) are plotted at 0 and 2.5 kyr in Fig. 3. The planet opens up a gap in the disc at around 2.5 kyr that is seen in the surface density profile at ~40 au. The density increase at the centre of this gap corresponds to the gas of the circumplanetary disc (see also the corresponding temperature increase at the same radius). The Toomre $Q$ parameter increases in the gap (due to the decrease in surface density), giving rise to a characteristic profile with two minima, one inside, and one outside the planet's orbit, which we refer to as $Q_{in}$ and $Q_{out}$, respectively. We see in Fig. 3 a that $Q_{out}$ is smaller than $Q_{in}$, whereas both are within 1–2, so that the gap edges are gravitationally unstable.

### 4.2 Binary evolution

For a binary mass ratio of $q_b = 1$ (see Fig. 4), the final binary separation is larger, apart from Run 3, in which it is slightly smaller (Fig. 4a). This behaviour is similar in the simulations with binary mass ratio $q_b = 0.25$ (see Fig. 5a). Binaries in Runs 6, 7, and 10 become wider with a stronger interaction with the disc, whereas in Run 8 and 9, the binary becomes tighter. Previous works of the evolution of binaries with circumbinary discs have reached conflicting results: the binary can either gain angular momentum and widen (Miranda, Muñoz & Lai 2017; Muñoz, Miranda & Lai 2019), or lose angular momentum and shrink (Heath & Nixon 2020; Tokovinin & Moe 2020), depending on the specific conditions of the disc and numerical considerations (Franchini et al. 2023). Heath & Nixon (2020) find that a thin circumbinary disc (e.g. $H/R \lesssim 0.1$) leads to the binary shrinking over time, whereas a thicker disc (like the ones we model here) leads to the widening of the binary.

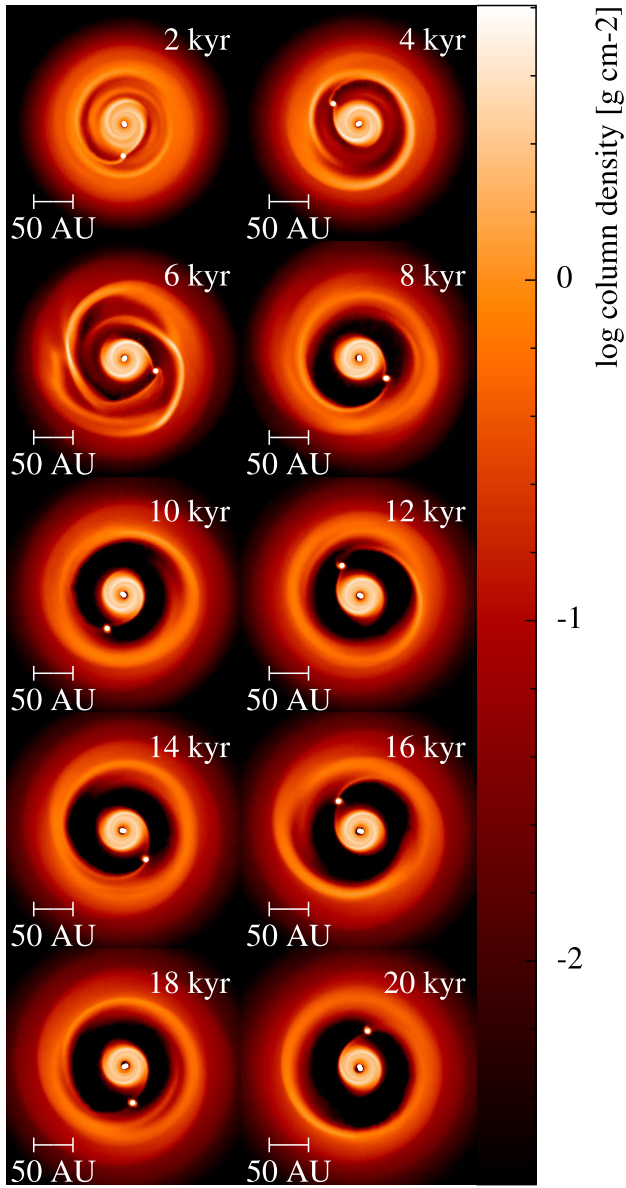

**Figure 1.** The evolution of the disc surface density (in g cm$^{-2}$) for Run 1 listed in Table 1. A 1 M$_J$ planet embedded at 50 au in a 0.1 M$_\odot$ disc, around a binary with mass ratio $q_b = 1$. The disc-planet interaction is shown from 2 kyr until the end of the simulation at 20 kyr.

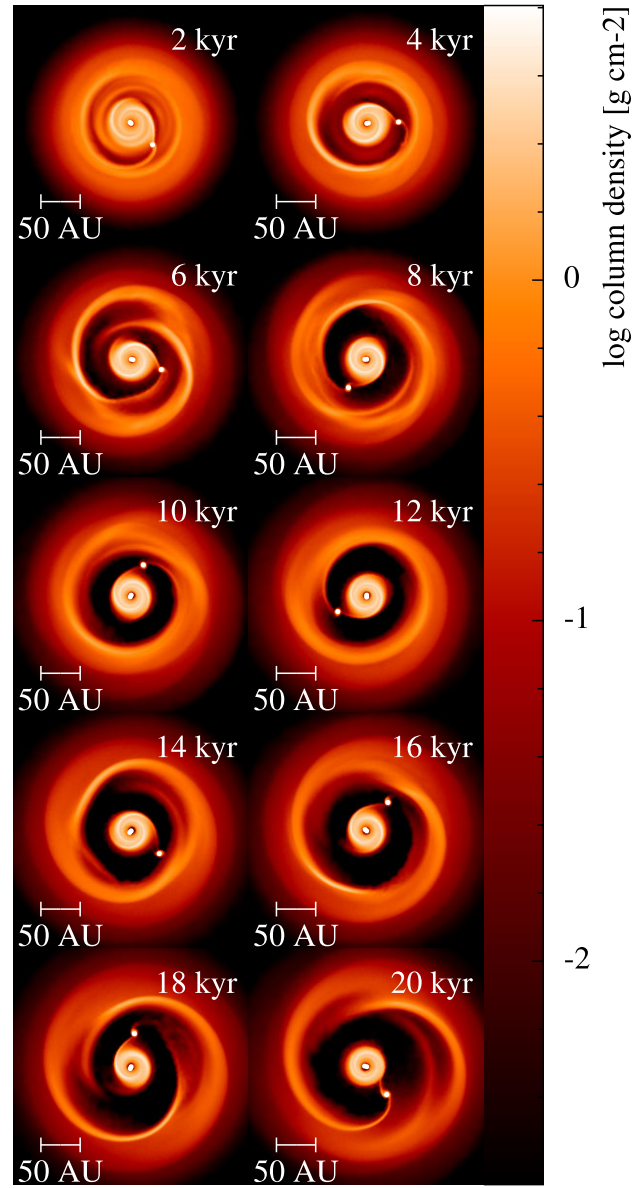

**Figure 2.** The evolution of the disc surface density (in g cm$^{-2}$) for Run 6 listed in Table 1. A 1 M$_J$ planet embedded at 50 au in a 0.1 M$_\odot$ disc, around a binary with mass ratio $q_b = 0.25$. The disc-planet interaction is shown from 2 kyr until the end of the simulation at 20 kyr.

The binary mass ratio in our simulations remains effectively the same for the systems with binary mass ratio $q_b = 1$, whereas for the uneven mass binary $q_b = 0.25$ there is a small increase as the secondary accretes more gas than the primary (see Figs 4b and 5b).

The eccentricity of the binary generally changes only by a small amount (see Figs 4c and 5c). However, for Run 2 and 7 (close binaries with separation 1 au and a high eccentricity of 0.2), the eccentricity drops to $e = 0$ within ~0.5 kyr (i.e. during the relaxation of the binary). This effective circularizing of the binary orbit makes these runs seemingly similar to Run 1 and Run 6, which have zero eccentricity. However, during the period of circularization the structure of the disc has been modified so that the embedded planet evolves differently.

### 4.3 Planet evolution

The evolution of the planet properties (semimajor axis, mass, and eccentricity) for all 10 runs and the benchmark run are shown in Figs 6 and 7.

#### 4.3.1 The role of the initial binary mass ratio

In the simulations with initial binary mass ratio $q_b = 1$ there is a larger range of final semimajor axes and final planet masses compared to simulations with initial binary mass ratio $q_b = 0.25$ (see Figs 6a,b and 7a,b). A higher binary mass ratio, $q_b = 1$, generally leads to farther outward migration compared to a low mass ratio, $q_b = 0.25$, binary. This is because the gravitational field of a binary with equal mass stars is more different than that of a single star, than the gravitational field of a binary with unequal mass stars, thus affecting the disc and the embedded planet more. This trend does not continue for the eccentricity of the planet (see Figs 6c and 7c) as in both set of simulations the range of final eccentricities are similar.

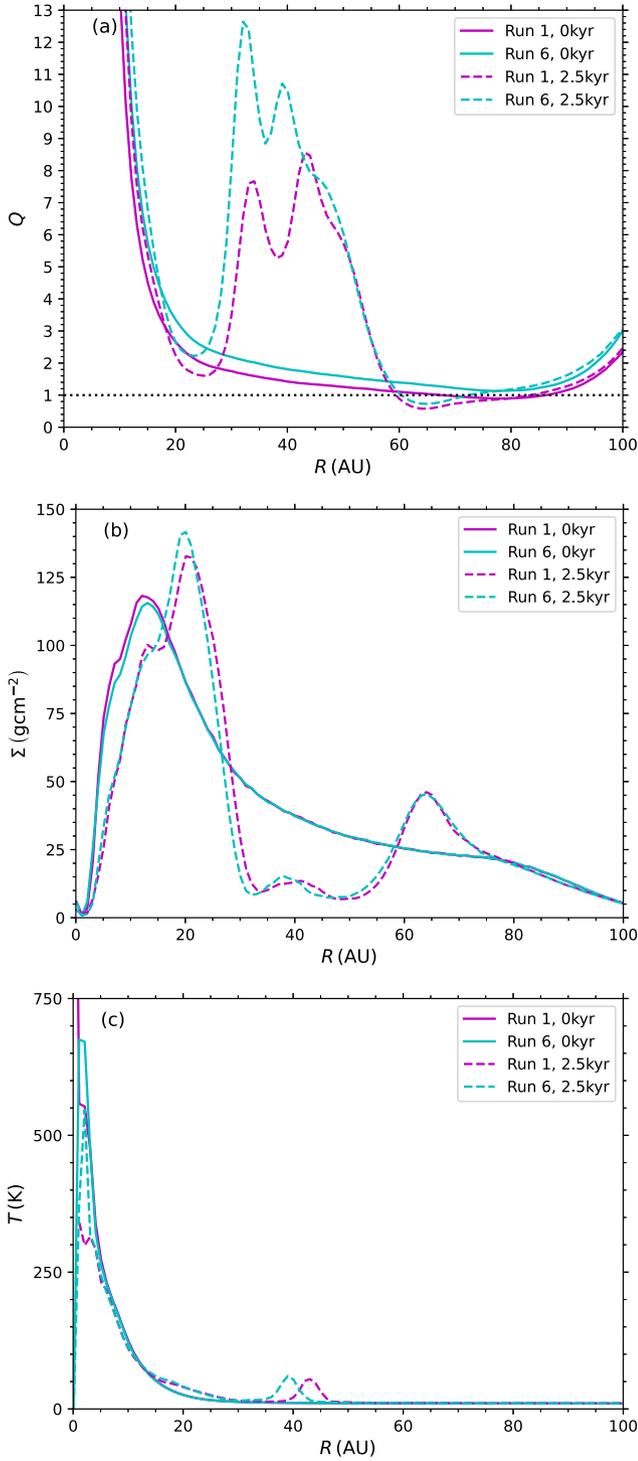

**Figure 3.** The Toomre parameter (a), the surface density (b) and the temperature profile (c), for Run 1 (magenta) and Run 6 (cyan) at $t = 0$ kyr (solid lines) and $t = 2.5$ kyr (dashed) plotted against the distance from the centre of mass of the binary.

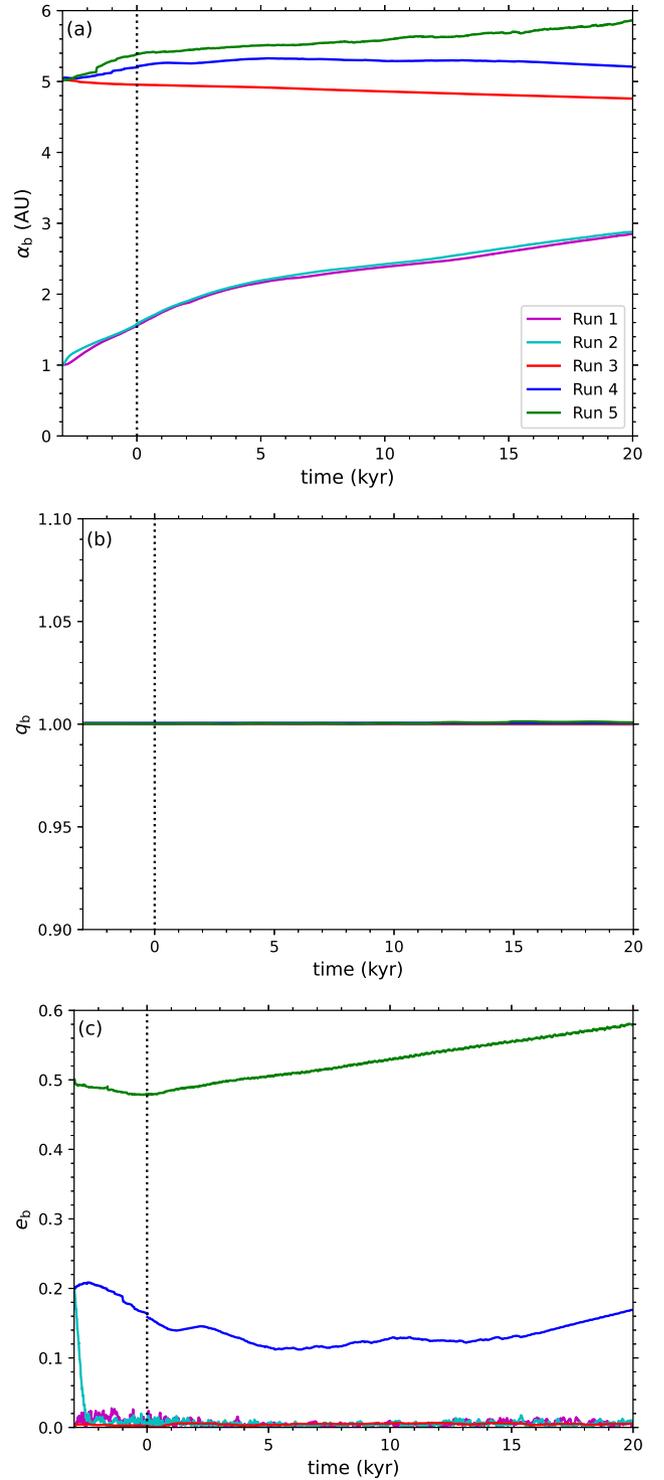

**Figure 4.** The evolution of the binary parameters in the simulations with a binary of initial mass ratio $q_b = 1$ (Runs 1 to 5, with colours as seen on the top graph). (a) The evolution the separation of the binary, (b) the evolution of the binary's mass ratio, and (c) the evolution of the binary's eccentricity. Negative times correspond to the period when the disc is allowed to relax (i.e. it is evolved before the planet is embedded in it).

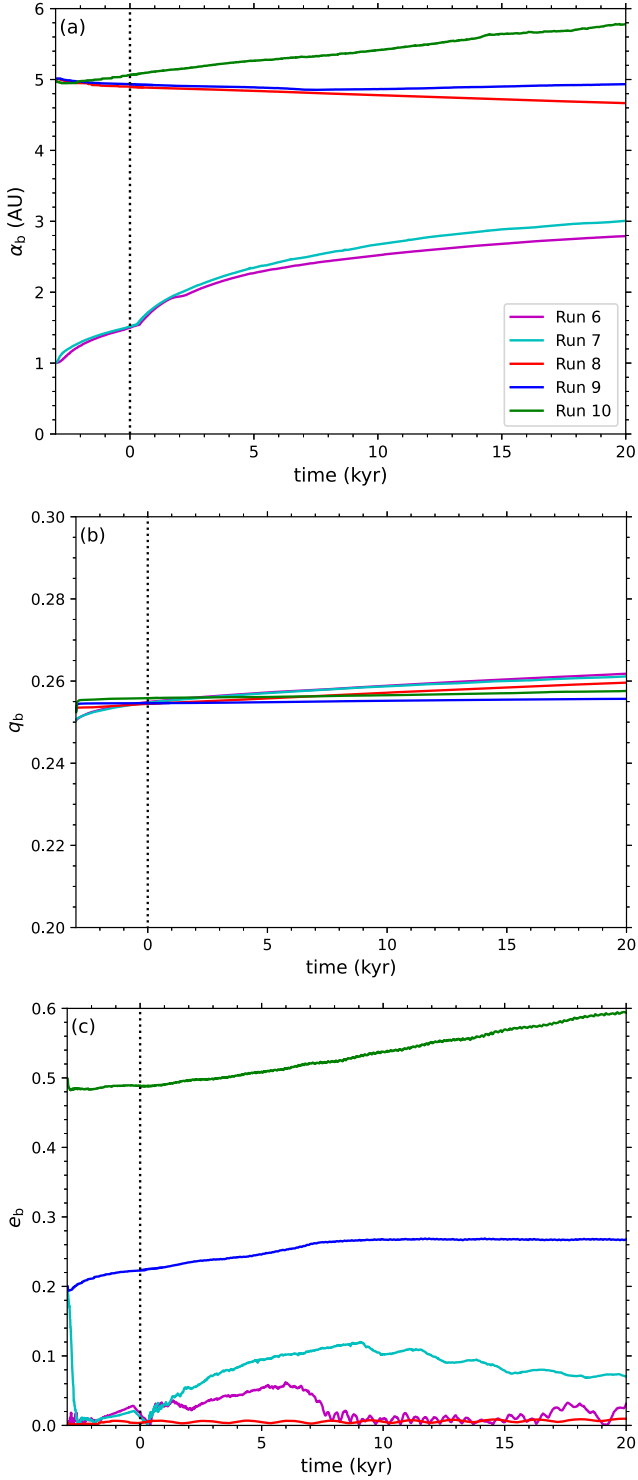

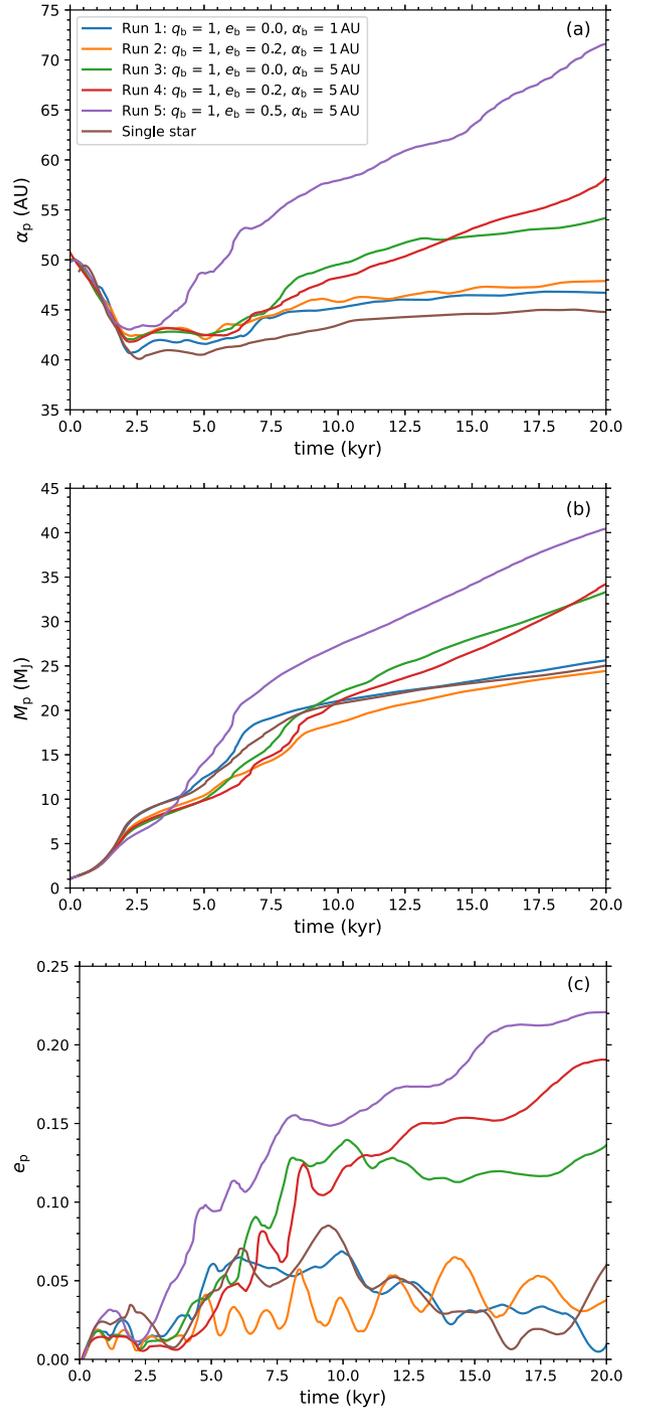

**Figure 5.** The evolution of the binary parameters in simulations with a binary of initial mass ratio $q_b = 0.25$ (Runs 6 to 10, with colours as seen on the top graph). (a) The evolution of the binary separation, (b) the evolution of the binary's mass ratio, and (c) the evolution of the binary's eccentricity. Negative times correspond to the period when the disc is allowed to relax.

**Figure 6.** The evolution of the $1\,M_J$ planet embedded in a circumbinary disc around a binary of initial mass ratio $q_b = 1$, with initial separation $\alpha_b = 1$ AU and $\alpha_b = 5$ au and initial eccentricity $e_b = 0.0$, 0.2, and 0.5 (as marked on the graph). (a) The evolution of the semimajor axis of the planet, (b) the evolution of the planet's mass, and (c) the evolution of the planet's eccentricity. For comparison, we also plot the evolution of a planet in a circumstellar disc around a single star with mass equal to the total mass of the binary (benchmark run; see Section 3.2).

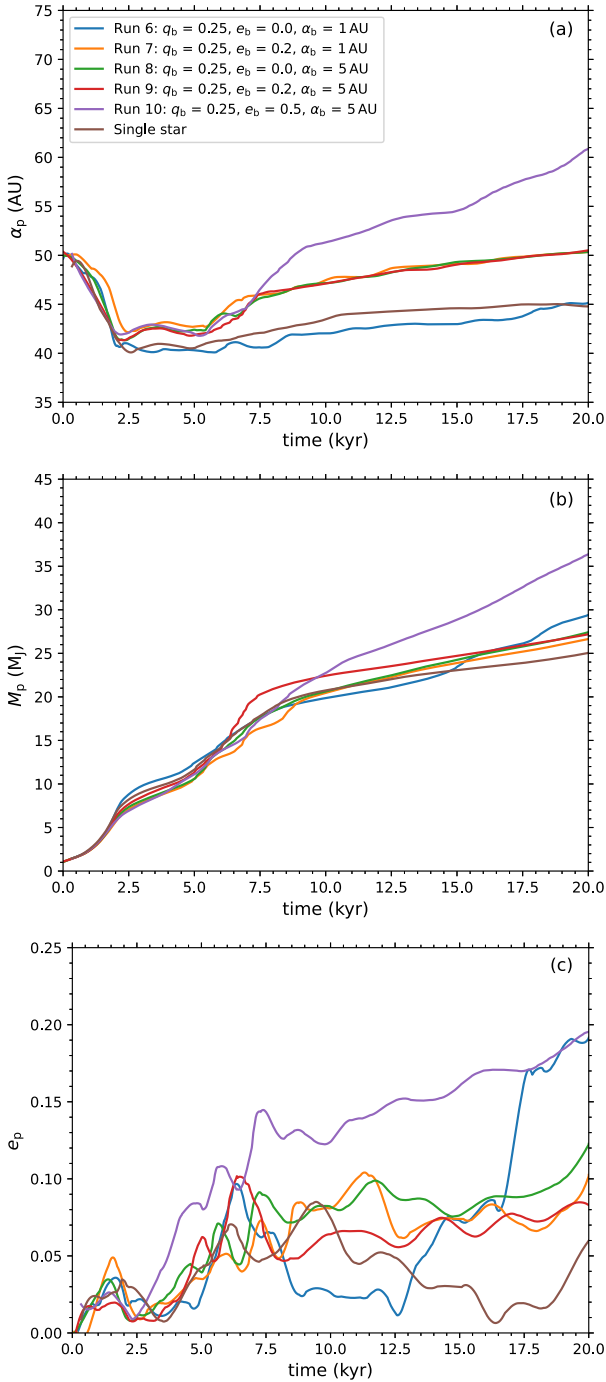

**Figure 7.** The evolution of the $1\,M_J$ planet embedded in a circumbinary disc around a binary of initial mass ratio $q_b = 0.25$, with initial separation $\alpha_b = 1$ au and $\alpha_b = 5$ au and initial eccentricity $e_b = 0.0$, 0.2, and 0.5 (as marked on the graph). (a) The evolution of the semimajor axis of the planet, (b) the evolution of the planet's mass, and (c) the evolution of the planet's eccentricity. For comparison, we also plot the evolution of a planet in a circumstellar disc around a single star with mass equal of the total mass of the binary (benchmark run; see Section 3.2.).

### 4.3.2 The role of the initial binary separation

In the simulations with initial binary separation of $\alpha_b = 1$ au the planet reaches a final semimajor axis between 45–50 au (see Figs 6a and 7a). Generally speaking the effect of a closer binary on planet migration is small compared to the effect of a wider binary. Larger initial binary separation of $\alpha_b = 5$ au leads to larger semimajor axes, between 45–72 au. This trend continues for the mass of the planet, with a larger initial binary separation leading to larger final masses (see Fig. 6b and 7b); for $\alpha_b = 1$ au, it is between $23 - 30\,M_J$ whereas for $\alpha_b = 5$ au, between $26 - 41\,M_J$. The evolution of the eccentricity of the planet follows a similar trend, with larger values for $\alpha_b = 5$ au than those with $\alpha_b = 1$ au (see Figs 6c and 7c). Generally speaking, larger initial binary separation leads to larger semimajor axis, mass, and eccentricity for the planet.

### 4.3.3 The role of the initial binary eccentricity

We find that increasing the initial eccentricity of the binary from $e_b = 0.0$ to 0.2 does not greatly affect the evolution of the planet's semimajor axis (see Fig. 6a and 7a). However, this is not the case when increasing this initial eccentricity to $e_b = 0.5$. This change leads to an accelerated outward migration, with the planet reaching much higher semimajor axis. A similar trend can be seen in the mass of the planet; an initial binary eccentricity of $e_b = 0.0$ and $e_b = 0.2$ leads to a similar planet mass evolution (see Figs 6b and 7b), where a more eccentric binary ($e_b = 0.5$) results in a larger planet mass. The planet's eccentricity follows a similar trend to that of the planet's mass, with the eccentric binary ($e_b = 0.5$) resulting to larger final planet eccentricity.

### 4.3.4 Comparison of the planet Hill radius with the planet sink radius

To adequately resolve the region around the planet, its sink radius needs to be much smaller than its Hill radius. The mass of the planet increases with time faster than the planet's semimajor axis decreases, while the mass of the binary does not change significantly. Therefore, the planet Hill radius (see equation 3) increases with time from its original value of $R_H = 3.4$ au, and it is always much larger than the planet sink radius, $R_{sink,p} = 0.1$ au. If we assume that the planet mass does not increase, then the minimum value of the Hill radius is obtained when the planet is closer to the star. Using typical values from the simulations ($a_p^{min} \sim 40$ au) and assuming $M_p \sim 1\,M_J$, we find a minimum possible value of $R_H^{min} \sim 0.6$ au (in reality the Hill radius is much larger, $\sim 6$ au, as the planet at this time has mass $M_p \sim 10\,M_J$).

## 5 DISCUSSION

The evolution of the planet is determined by the disc in which it forms. The structure of this circumbinary disc and how it changes as it interacts with the binary, is key to understanding the planet's evolution.

### 5.1 Migration time-scale and velocity

Fig. 8 shows how the migration time-scale and the migration velocity of the planet evolves with time for Runs 1–10 and the single star simulation. Fig. 8a shows that the planet initially has a short migration time-scale, $< 1$–2 kyr, during the Type I phase of its migration while it opens up the gap in the disc. This increases significantly once the gap opens up in the disc. Eventually the planet starts to migrate outwards with a rather long migration time-scale (initially around $\sim 5$–20 kyr). This is a similar pattern to the one of a planet migrating in a circumstellar disc (benchmark run).

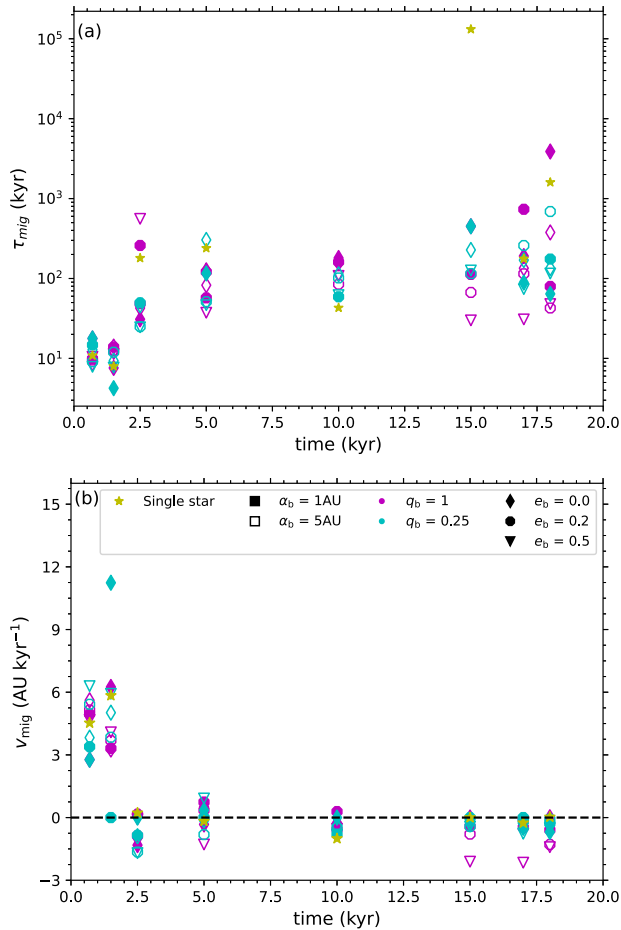

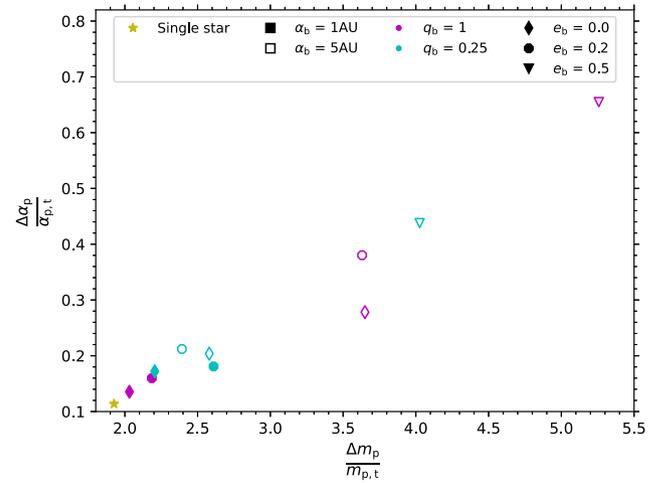

**Figure 9.** The fractional change in the mass of the planet plotted against the fractional change in the semimajor axis of the planet, between the time the planet opens up a gap in the disc (2.5 kyr) and the end of the simulation (20 kyr).

(where $m_{p,f}$ is the mass of the planet at $t = 20$ kyr and $m_{p,t}$ is the mass of the planet at $t = 2.5$ kyr).

We see that the larger the change of the planet's semimajor axis the larger its mass growth. This shows that planet migration plays an important role in the evolution of the mass of the planet. It is also evident that the presence of a binary instead of a single star of the same mass as the binary (i.e. a circumbinary instead of a circumstellar disc) enhances both planet migration and planet mass growth. A binary with equal mass stars generally promotes planet migration compared to a binary with unequal mass stars (magenta versus cyan points in Fig. 9). A wider binary ($a_b = 5$ au) enhances planet migration more than a close binary ($a_b = 1$ au; open symbols versus filled in symbols). More eccentric binaries also enhance planet migration (circles versus triangles; Fig. 9), with this enhancement being more prominent for the high eccentricity case (see discussion in Section 4.3.3). Finally, the binary mass ratio does not seem to have a strong effect on planet migration.

## 5.3 The connection between the gap edges and planet migration

Fig. 10a shows the minimum of the Toomre parameter, $Q_{in}$, at the inner edge of the gap opened up by the planet (see 4.1) plotted against the fractional change of the planet's semimajor axis between the time that the planet opens up a gap and the end of the simulation (equation 6). Fig. 10b shows a similar plot for the minimum Toomre parameter at the outer edge of the gap, $Q_{out}$.

It is evident from these graphs that a binary with unequal mass components ($q_b = 0.25$) has a smaller effect on the disc structure than a binary with equal mass components, as both $Q_{in}$ and $Q_{out}$ are lower in the latter case.

We see (Fig. 10a) that for the wide binary ($a_b = 5$ au; open symbols) the outward migration of the planet is farther out when the inner minimum Toomre parameter $Q_{in}$ is lower, i.e. when the gap inner edge is more unstable. For the close binary ($a_b = 1$ au) there seems to be no such correlation, with an approximate similar semimajor axis fractional change for all these runs. Similarly, there is no correlation between the outer minimum Toomre parameter $Q_{out}$ and migration (see Fig. 10b).

**Figure 8.** The migration time-scale (a), and migration velocity (b), at different times for all runs. Colours correspond to different binary mass ratios: magenta indicates $q_b = 1$, and cyan $q_b = 0.25$. Symbols correspond to the initial eccentricity of the binary: diamond to $e_b = 0.0$, an octagon to $e_b = 0.2$, and triangle to $e_b = 0.5$. Finally, filled symbols correspond to an initial binary separation of $a_b = 1$ au and unfilled symbols to an initial separation of $a_b = 5$ au. Negative values to the migration velocity correspond to outward migration. For comparison, we also plot the migration time-scale and velocity for a planet in a circumstellar disc around a single star with mass equal to the total mass of the binary (benchmark run; see Section 3.2.).

The migration velocity is initially high as the planet is opening up a gap but once the gap is opened up the migration velocity becomes smaller in magnitude and negative (i.e. in the outward direction).

## 5.2 The connection between planet migration and planet growth

Fig. 9 shows the fractional change of the planet's semimajor axis between the time that the planet opens up a gap and the end of the simulation, i.e.

$$\frac{\Delta \alpha_p}{\alpha_{p,t}} = \frac{\alpha_{p,f} - \alpha_{p,t}}{\alpha_{p,t}},$$ (6)

(where $\alpha_{p,f}$ is the semimajor axis of the planet at $t = 20$ kyr and $\alpha_{p,t}$ is the semimajor axis of the planet at $t = 2.5$ kyr), plotted against the fractional change of the planet's mass during the same period,

$$\frac{\Delta m_p}{m_{p,t}} = \frac{m_{p,f} - m_{p,t}}{m_{p,t}},$$ (7)

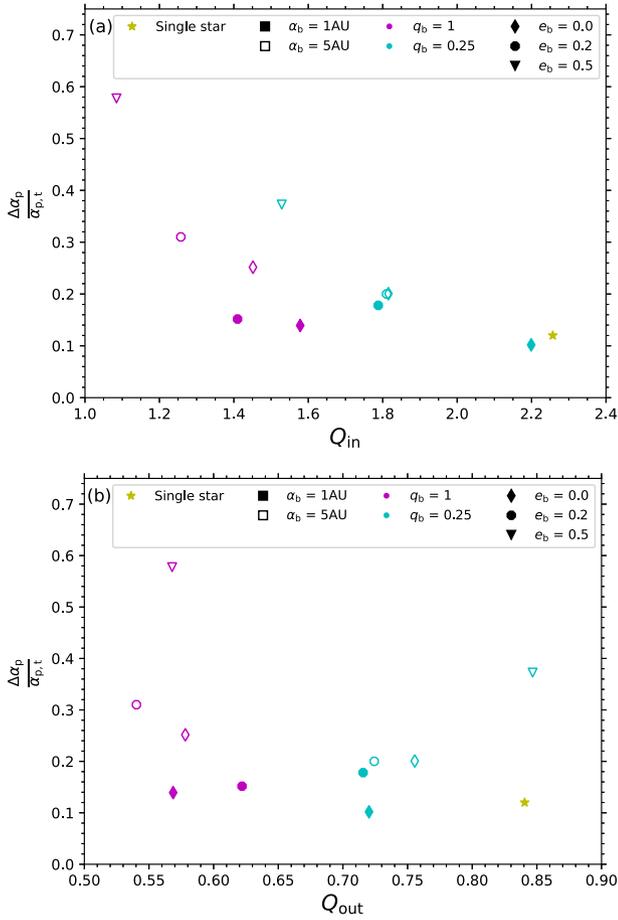

**Figure 10.** The minimum of the Toomre parameter at the inner edge of the gap (a), and the outer edge of the gap (b), plotted against the fractional change in the semimajor axis of the planet at the time between the gap opening in the disc (2.5 kyr) and the end of the simulation (20 kyr).

### 5.4 The connection between binary evolution and planet migration

Fig. 11 shows the fractional change of the planet's semimajor axis $\frac{\Delta\alpha_p}{\alpha_{p,t}}$ (see equation 6), plotted against the fractional change of the binary separation $\frac{\Delta\alpha_b}{\alpha_{b,t}}$, during the same interval. We see no correlation between the change in the semimajor axis of the planet and the change in the semimajor axis of the binary, which suggests that the binary does not affect directly the planet's evolution, but indirectly through modifying the environment in which the planet evolves, i.e. the circumbinary disc.

### 6 CONCLUSIONS

We used the SPH code SEREN to study the migration of a wide-orbit Jupiter-mass planet (semimajor axis 100 au) in a massive circumbinary disc $(0.1\,\mathrm{M}_\odot)$ that is marginally unstable (but not fragmenting). We first performed a simulation of a Jupiter-mass planet evolving in a circumstellar disc as in Stamatellos & Inutsuka (2018). Then, we performed several simulations of a Jupiter-mass planet evolving in a circumbinary disc around a binary with the same total mass $(1\,\mathrm{M}_\odot)$ as the single star in the circumstellar simulation. We varied the binary parameters to see their effect on the evolution of the planet embedded within the circumbinary disc. Our aim was to see how the presence of a binary may affect planet migration and to investigate whether a planet formed on a wide orbit (e.g. through gravitational instability) could survive on a wide orbit and not migrate inwards.

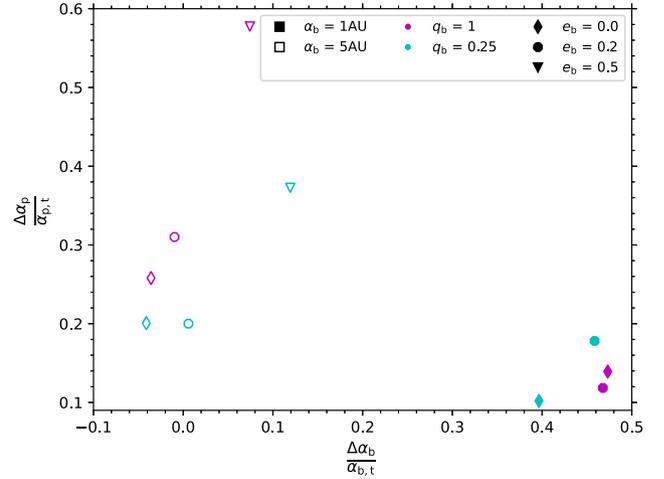

**Figure 11.** The fractional change in the binary separation plotted against the fractional change in the semimajor axis of the planet (see discussion the text) between the time between the planet opening the gap in the disc (2.5 kyr) and the end of the simulation (20 kyr).

We find that a Jupiter-mass planet, embedded in a circumbinary disc, initially migrates inwards but it is able to open up a gap in the disc, and then migrates outwards. Our results are in agreement with previous similar studies (Stamatellos 2015; Stamatellos & Inutsuka 2018). The migration pattern seen in our simulations is also similar to that seen in Pierens & Nelson (2008), Kley & Haghighipour (2015), and Kley, Thun & Penzlin (2019), albeit in a different regime (they study low-mass circumbinary discs with planets close to the binary, so that that resonances between the planet and the binary may be important).

Our main result is that the presence of a binary instead of a single star (i.e. a circumbinary rather than in a circumstellar disc) makes outward planet migration faster, resulting in a faster mass growth for the planet. A binary with equal mass stars helps the planet to migrate farther out compared to a binary with unequal mass stars. A larger binary separation enhances planet migration and mass growth, and the same holds for a larger binary eccentricity. The binary affects indirectly the evolution of the planet by altering the structure of the circumbinary disc. We conclude that wide-orbit giant planets attain wider final orbits due to migration around binary stars than around single stars.

A possible system that this predicted outward migration may be relevant is Delorme 1(AB)b (Delorme et al. 2013), a planetary-mass companion $(12{-}14\,\mathrm{M}_J)$ orbiting at a projected separation of ∼84 au from a pair of young M-dwarfs with masses 0.19 and $0.17\,\mathrm{M}_\odot$, and projected separation 12 au. This planet shows signs of accretion as evidenced by strong emission in H$_\alpha$ and other lines in the near-infrared (Eriksson et al. 2020; Betti et al. 2022), indicating interaction with a disc. According to the models we presented here, it may have formed closer to the binary pair and then migrated outwards to the position observed now, while significantly grown in mass.

Extending this work at different binary mass ratios, eccentricities, separations, total binary mass, disc mass and properties, and different-mass exoplanets would be useful in gaining a deeper understanding as to how young planets evolve in binary systems.

ACKNOWLEDGEMENTS

We thank the anonymous referee for his/her constructive review that helped improving the paper. The simulations were performed using the UCLan High Performance Computing (HPC) facility. This research has made use of the NASA Exoplanet Archive, which is operated by the California Institute of Technology, under contract with the National Aeronautics and Space Administration under the Exoplanet Exploration Program. We thank David Hubber for the development of SEREN. Surface density plots were produced using SPLASH (Price 2007).

DATA AVAILABILITY

The simulation data used for this paper can be provided by contacting the authors.

This paper has been typeset from a TEX/LATEX file prepared by the author.